\documentclass[12pt]{article}

\usepackage{afterpage,amsfonts,amsmath,amssymb,array}
\usepackage{epsfig,hhline,longtable}
\usepackage{tabularx}
\usepackage{enumerate}
 \def\dir{}

\usepackage{\dir chicago}
\usepackage{\dir klett}
\usepackage{\dir def_the}

\def\secref{Section~\ref}
\def\bE{{{\mbox {\boldmath $E$}}}}
\def\tucF{\widetilde{\ucF}}

\newcommand{\Figgnunu}[3]{\begin{figure}[ht]
\centering
 \vspace*{-3mm}
 \epsfig{file=#1, width=\fwnu}
 \vspace*{-3mm}
 \caption{#2.}
 \label{#3}
\end{figure}}

\def\fwnu{10cm} %{11cm} % for normal gnuplot pictures UTS

\begin{document}
\thispagestyle{empty}

\begin{center}
{\LARGE \bf On Honest Times in Financial Modeling }
\end{center}

\vspace*{1.5cm}
\begin{center}

{\large
\renewcommand{\thefootnote}{\arabic{footnote}}
{\bf Ashkan Nikeghbali}\footnote{Institut f\"ur Mathematik,
 Universit\"at Z\"urich, Winterthurerstrasse 190,
 CH-8057 Z\"urich, \\
\hspace*{0.6cm} Switzerland} and
 \renewcommand{\thefootnote}{\arabic{footnote}}
{\bf Eckhard Platen}\footnote{University of Technology Sydney,
School of Finance $\&$ Economics and Department of \\
\hspace*{0.6cm} Mathematical Sciences, PO Box 123, Broadway, NSW,
2007, Australia }

}
\vspace*{1.0cm}

\today

\end{center}

{\small
\begin{center}
\vspace*{1cm}
\begin{minipage}[t]{13cm}
{\bf Abstract.} This paper demonstrates the usefulness and importance of the concept of honest times to financial modeling. It studies a financial market with asset
prices that follow jump-diffusions with negative jumps. The
central building block of the market model is its growth optimal
portfolio (GOP), which maximizes the growth rate of strictly
positive portfolios. Primary security account prices, when
expressed in units of the GOP, turn out to be nonnegative local
martingales. In the proposed framework an equivalent risk neutral
probability measure need not exist. Derivative prices are obtained
as conditional expectations of corresponding future payoffs, with
the GOP as numeraire and the real world probability as pricing
measure. The time when the global maximum of
a portfolio with no positive jumps, when  expressed in units of the GOP, is reached,  is shown to be a generic representation of an honest time.   We provide a
general formula for the law of such honest times and  compute the conditional distributions of the global maximum of a portfolio in this framework. Moreover, we provide a stochastic integral representation for  uniformly integrable martingales whose terminal values are  functions of the global maximum of a portfolio. These formulae are model independent and universal. We also specialize our results to some examples where we hedge a payoff that arrives at an honest time.
 \\[1cm]
\end{minipage}
\end{center}}


\newpage
\section{Introduction}
In this paper we consider a general class of jump-diffusion
financial market models under the benchmark approach, described in
\citeN{PlatenHe06}. Our main goal is, within this framework,  to
provide modelers and investors with extra tools based on the
concept of honest times, which are random times that are not
stopping times and that are somehow hidden. We show that
quantities of interest, such as  the law  of the  time when the
global maximum value of a portfolio is reached, or the conditional
laws of the global maximum value of a portfolio, can be computed
explicitly. Most importantly, these results will be very robust
since they  do not depend on the specifications of the underlying
model. Surprisingly, these results reveal the universal feature
that no Markovian setting needs to be assumed. This can be
achieved thanks to the use of  martingale techniques.

More precisely, we let security prices follow {\em jump
diffusions}. There exists a range of literature on modeling and
pricing for jump diffusions, starting with \citeN{Merton76}. For a
detailed discussion of this area the reader is referred to
\citeN{ContTa04}. Different to most authors we will avoid the
standard assumption on the existence of an equivalent risk neutral
probability measure. In this way important freedom is gained for
financial modeling. All tasks of portfolio optimization,
derivative pricing, hedging  and risk management can still be
consistently performed, see \citeN{PlatenHe06}. The {\em growth
optimal portfolio}, which maximizes the growth rate of all
strictly positive portfolios is the central building block of the
market model. It is also the numeraire portfolio in the sense of
\citeN{Long90} and \citeN{Becherer01}. When used as numeraire or
benchmark, it makes all benchmarked portfolios  local martingales
and, thus, all nonnegative benchmarked portfolios
supermartingales. This supermartingale property excludes
automatically a strong form of arbitrage. Furthermore, in a
complete market, nonnegative replicating portfolios, when
expressed in units of the growth optimal portfolio are minimal
when they form martingales. Therefore, benchmarked derivative
prices will be obtained as martingales.

An {\em honest time\/} is by definition the end of an optional
set, see for example \citeN{Jeulin80}, \citeN{DellacherieMaMe92}
for  references and details. For instance,  the last time when the
maximum of some benchmarked nonnegative portfolio with no positive
jumps is reached is an example of an honest time. It is a random
time but not a stopping time, which makes its analysis more
delicate. Honest times have been intensively studied in stochastic
analysis, see e.g. \citeN{Chung73},  where they play an important
role in the theory of enlargements of filtrations, see
\citeN{Barlow78}, \citeN{Jeulin80}, \citeN{JeulinYo85},
\citeN{Yor97} and \citeN{NikeghbaliYo06}, in the characterizations
of strong Brownian filtrations, \shortciteN{Barlow+98},
\citeN{MansuyYo06}, and in path decompositions of diffusions, see
\citeN{Jeulin80}, \citeN{Salminen85}, \citeN{Salminen97}, and \citeN{Nikeghbali06e}. Honest times have also
recently received some attention in mathematical finance, e.g. for
modeling default in \citeN{ElliottJeYo00}, for insider trading in
\citeN{Imkeller02} and  for pricing options in the Black-Scholes
framework in \citeANP{MadanRoYo08b}
\citeyear{MadanRoYo08b,MadanRoYo08o,MadanRoYo08u}. In this paper,
we shall pursue these last trends in mathematical finance and show
 that honest times can serve as rather
useful and important random quantities in financial modeling.

In some special cases, such as the time of the last zero before
time one of a standard Brownian motion or some Bessel process or
the laws of last passage times of transient diffusions, the law of
an honest time can be explicitly characterized, see
\citeN{Levy39}, \citeN{PitmanYo81}, \citeN{BorodinSa96},
\citeN{BarlowPiYo89} and \citeANP{Nikeghbali06a}
\citeyear{Nikeghbali06a,Nikeghbali06s} for examples. We will rely
in this paper on a characterization of honest times given in
\citeN{NikeghbaliYo06} and take this to be our basic reference
without further mentioning. The important fact that nonnegative
benchmarked portfolios form local martingales in jump diffusion
markets will play a crucial role. In this context honest times
related to the last maxima of benchmarked securities will be
studied. These particular random times are extremely interesting
from an investor's point of view. They can describe, for instance,
the time for the highest value ever of the security relative to
the benchmark. The law of this time is valuable information for an
investor. We will provide a general formula for the law of an
honest time (Theorem \ref{the:law5.2}). We then specialize it to
the last time when a benchmarked nonnegative portfolio reaches its
maximum. We shall also provide  the conditional distributions of
the global maximum of a benchmarked nonnegative portfolio.
Moreover, we give a stochastic integral representation for any
martingale whose terminal value is a function of the global
maximum of a benchmarked nonnegative portfolio. This suggests
certain hedging strategies for reaching this payoff which arises
at some honest time.

The structure of the paper is as follows. In \secref{sec:law2} we
describe the underlying general jump diffusion market. In
\secref{sec:law4} important facts on honest times are given and a
general formula for their laws is derived. \secref{sec:law5}
provides some examples concerning the actual computation of such
laws.  For notations and definitions that are
used but not explained in the paper we refer to \citeN{RevuzYo99}
or \citeN{Protter90}.

\section{Jump Diffusion Market}
 \setcA
 \label{sec:law2}
We consider a market where continuously evolving risk is modeled
by $m$ independent standard Wie\-ner processes $\tW^k=\{\tW^k_t,\,
\tgo\}$, $\,k \in \cNm$, $m \in \cNd$, $d \in \cNs$, defined on a
filtered probability space $(\Omega,\cF,\ucF,\bP)$. We also
consider events of certain types, for instance, corporate
defaults, operational failures or catastrophic events that are
reflected in traded securities. Events of the $k$th type shall be
counted by the adapted $k$th {\em counting process\/}
$p^k=\{p^k_t,\,\tgo\}$, whose {\em intensity\/}
$h^k=\{h^k_t,\,\tgo\}$ is a given predictable, strictly positive
process with
 \BE
 \label{ca2.2}
 \int^T_0 h^k_s \,ds < \infty,
 \end{equation}
almost surely for $\tgo$ and $k \in \{m+1,\ldots,d\}$.
Furthermore, we introduce the $k$th {\em normalized jump
martingale\/} $q^k=\{q^k_t,\,\tgo\}$ with stochastic differential
 \BE
 \label{ca2.3}
 dq^k_t = \l(dp^k_t-h^k_t\,dt\r)\l(h^k_t\r)^{-\frac12}
 \end{equation}
for $k \in \{m\!+\!1,\ldots,d\}$ and $\tgo$, which represents the
k-th source of {\em event driven risk}. We not only compensate but
also normalize the above sources of event risk to make these
comparable with the previously introduced standard Wiener
processes which provide the sources of continuous risk. It is
assumed that the above jump martingales do not jump at the same
time.

The evolution of traded risk is then modeled by the vector process
of independent martingales $\bW= \{
\bW_t=(W^1_t,\ldots,W^d_t)^\top,\,\tgo\}$, where $W^1_t=\tW^1_t$,
\ldots, $W^m_t=\tW^m_t$ are the above Wiener processes, while
$W^{m+1}_t=q^{m+1}_t$, \ldots, $W^d_t=q^d_t$ represent compensated
and normalized counting processes. The filtration
$\ucF=(\cF_t)_{\tgo}$ is assumed to be the augmentation under
$\bP$ of the natural filtration $\cF^W$, generated by the vector
process $\bW$. This filtration  satisfies the usual conditions and $\cF_0$ is
the trivial $\sigma$-algebra. Note that the conditional variance
of the $k$th source of traded risk equals
 \BE
 \label{ca2.4}
 \bE\l(\l(W^k_{t+h} -W^k_t\r)^2\sB|\cF_t\r) = h
 \EE
for all $\tgo$, $k \in \cNd$ and $h >0$.

For the securitization of the $d$ sources of traded risk, we
introduce $d$ {\em risky primary security accounts}, whose values
at time $t$ are denoted by $X^j_t$, for $j \in \cNd$. Each of
these accounts contains shares of one kind with all dividends
reinvested. Furthermore,  the 0th primary security
account $X^0 = \{X^0_t,\, \tgo\}$,  is the locally riskless
{\em savings account} that continuously accrues the short term
interest rate $r_t$. We assume that the nonnegative $j$th primary
security account value $X^j_t$ at time $\tgo$ satisfies the
stochastic differential equation (SDE)
 \BE
 \label{ca2.5}
 dX^j_t = X^j_{t-} \l(a^j_t\,dt + \sum^d_{k=1}
 b^{j,k}_t\,dW^k_t \r)
 \end{equation}
with initial value $X^j_0>0$, $j \in \cNod$. Since $X^0_t$ models
the savings account, we have $a^0_t = r_t$ and $b^{0,k}_t = 0$ for
$\tgo$ and $k \in \cNd$. We assume that the processes $r$, $a^j$,
$b^{j,k}$ and $h^k$ are finite and predictable, and such that a
unique strong solution for the system of SDEs \eqref{ca2.5}
exists. To guarantee strict positivity for each primary security
account we assume
 \BE
 \label{ca2.7'}
 b^{j,k}_t > - \sqrt{h^k_t}
 \EE
for all $\tgo$, $\,j \in \cNd$ and $k \in \{m+1,m+2,\ldots,d\}$.
Furthermore, we make the following assumption.

 \begin{ass}
 \label{ass:ca2.1} \quad
The {\em generalized volatility matrix\/}
$\bb_t=[b^{j,k}_t]^d_{j,k=1}$ is {\em invertible\/} for every
$\tgo$, and allows only downward jumps.
 \end{ass}

The invertibility of the generalized volatility matrix provides
the unique link between the sources of traded risk and the primary
security accounts. Negative jumps in equities are the most
important ones to model. These are caused, for instance, by
defaults and catastrophic events. Therefore, we can focus in our
analysis of honest times on models where there are no positive
jumps in primary security accounts. This assumption fits well into
the concept of honest times,  see \citeN{NikeghbaliYo06}.
Assumption~\ref{ass:ca2.1} allows us to introduce the {\em market
price of risk\/} vector
 \BE
 \label{ca2.8}
 \vtheta_t = (\theta^1_t,\ldots,\theta^d_t)^\top =
 \bb^{-1}_t\,[\ba_t-r_t\,\b1 ]
 \end{equation}
for $\tgo$. Here $\ba_t=(a^1_t,\ldots,a^d_t)^\top$ is the {\em
appreciation rate vector\/} and $\b1=(1,\ldots,1)^\top$ the {\em
unit vector\/}. Using \eqref{ca2.8}, we can rewrite the SDE
\eqref{ca2.5} in the form
 \BE
 \label{ca2.9}
 dX^j_t
 = X^j_{t-} \l(r_t\,dt + \sum^d_{k=1} b^{j,k}_t\,(\theta^k_t\,dt +
 dW^k_t) \r)
 \end{equation}
for $\tgo$ and $j \in \cNod$. For $k \in \cNm$, the quantity
$\theta^k_t$ expresses the {\em market price of risk\/} with
respect to the $k$th Wiener process $W^k$, and for $k \in
\{m+1,\ldots,d\}$, it can be interpreted as the {\em market price
of $k$th event risk}.

The vector process $\bX=\{\bX_t=(X^0_t,\ldots,X^d_t)^\top,$
$\tgo\}$ characterizes the evolution of all primary security
accounts. We say that a predictable stochastic process
$\vdelta=\{\vdelta_t=(\delta^0_t,\ldots,\delta^d_t)^\top$,
$\tgo\}$ is a {\em strategy} if it is $\bX$-integrable. The $j$th
component of $\vdelta$ denotes the number of units of the $j$th
primary security account held at time $\tgo$ in a portfolio, $j
\in \cNod$. For a strategy $\vdelta$ we denote by $X^\delta_t$ the
value of the corresponding portfolio process at time $t$, when
measured in units of the domestic currency. Thus, we set
 \BE
 \label{ca2.10}
 X^\delta_t = \sum^d_{j=0} \delta^j_t\,X^j_t
 \end{equation}
for $\tgo$. A strategy $\vdelta$ and the corresponding portfolio
process $X^\delta = \{X^\delta_t$, $\tgo\}$ are called {\em
self-financing\/} if
 \BE
 \label{ca2.11}
 dX^\delta_t = \sum^d_{j=0} \delta^j_t\,dX^j_t
 \end{equation}
for all $\tgo$. In what follows we will only consider
self-financing portfolios.

For a given strategy $\vdelta$, generating a strictly positive
portfolio process $X^\delta = \{X^\delta_t, \tgo\}$, let
$\pi^j_{\delta,t}$ denote the {\em fraction\/} of wealth that is
invested in the $j$th primary security account at time $t$, that
is,
  \BE
  \label{ca2.12}
  \pi^j_{\delta,t} = \delta^j_t\,\frac{X^j_t}{X^\delta_t}
  \end{equation}
for $\tgo$ and $j \in \cNod$. By \eqref{ca2.10} these fractions
always add to one. In terms of the vector of fractions
$\vpi_{\delta,t}= (\pi^1_{\delta,t},\ldots,$
$\pi^d_{\delta,t})^\top$ we obtain for $X^\delta_t$ from
\eqref{ca2.11}, \eqref{ca2.9} and \eqref{ca2.12} the SDE
 \BE
 \label{ca2.14}
 dX^\delta_t
 = X^\delta_{t-}\! \left\{r_t\,dt + \vpi_{\delta,t-}^\top\,\bb_t\,(\vtheta_t\,dt
 + d\bW_t) \right\}
\end{equation}
for $\tgo$. The following assumption ensures that no strictly
positive portfolio explodes in our market.

\begin{ass}
 \label{ass:ca2.3}
\quad We assume that
 \BE
 \label{ca2.18}
 \sqrt{h^k_t} > \theta^k_t
 \end{equation}
for all $\tgo$ and $k \in \{m+1,\ldots,d\}$.
 \end{ass}

Following \citeN{PlatenHe06}, this allows us to introduce for the
given jump diffusion market the growth optimal portfolio (GOP)
$X^\deltas$, which maximizes  expected logarithmic utility and, thus, the
growth rate of strictly positive portfolios. It satisfies the SDE
 \BA
 \label{ca2.21}
 dX^\deltas_t &=& X^\deltas_{t-} \l(r_t\,dt + \sum^m_{k=1}
 \theta^k_t\l(\theta^k_t\,dt +dW^k_t\r) \right. \n2
 && \left. \qquad +\, \sum^d_{k=m+1}
 \frac{\theta^k_t}{1-\theta^k_t\,(h^k_t)^{-\frac12}} \l(\theta^k_t\,dt
 + dW^k_t\r) \r)
  \end{eqnarray}
for $\tgo$, with $X^\deltas_0 >0$. This portfolio is also the
numeraire portfolio in the sense of \citeN{Long90} and
\citeN{Becherer01}, and is  in several mathematical manifestations
the best performing portfolio. We use $X^\deltas$ as benchmark,
and accordingly,  call prices, when expressed in units of
$X^\deltas$, {\em benchmarked prices}. By the It\^o formula,
\eqref{ca2.14} and \eqref{ca2.21}, a {\em benchmarked portfolio
process\/} $N^\delta=\{N^\delta_t, \tgo\}$, with
 \BE
 \label{ca2.26}
 N^\delta_t = \frac{X^\delta_t}{X^\deltas_t}
 \end{equation}
 for $\tgo$, satisfies the SDE
 \begin{eqnarray}
 \label{ca2.30}
 dN^\delta_t \sq N^\delta_{t-} \l(\sum^m_{k=1}
 \left\{ \sum^d_{j=1} \pi^j_{\delta,t}\, b^{j,k}_t-\theta^k_t \right\} dW^k_t
 \right.\n2
 \sl \left.+\sum^d_{k=m+1}\! \! \left\{ \l(\sum^d_{j=1}
 \pi^j_{\delta,t-} \,b^{j,k}_t \r) \!
 \l(1-\frac{\theta^k_t}{\sqrt{h^k_{t}}}\r)
 - \theta^k_t \right\} dW^k_t\r) \n2
 \sq - N^\delta_{t-}\,\sum^d_{k=1} \sum^d_{j=0}
 \pi^j_{\delta,t-}\,\sigma^{j,k}_t\,dW^k_t
 \end{eqnarray}
for $\tgo$. In this  equation we wrote $\sigma^{0,k}_t$ instead
of $\theta^k_t$, for $k \in \cNd$, and used the notation
 \BE
 \label{ca2.29}
 \sigma^{j,k}_t = \left\{ \begin{array}{c@{\quad \mbox{for} \quad}l}
 \sigma^{0,k}_t - b^{j,k}_t &
 k \in \cNm \\[1.5ex]
 \sigma^{0,k}_t - b^{j,k}_t
 \l(1-\frac{\sigma^{0,k}_t}{\sqrt{h^k_t}}\r)
  & k \in \{m+1,\ldots,d\}
  \end{array} \right.
 \end{equation}
for $j \in \cNd$, $\tgo$. The SDE~\eqref{ca2.30} shows that the
dynamics of any benchmarked portfolio is driftless. Thus, a
nonnegative benchmarked portfolio $N^\delta$ forms a local
martingale. This also means that a benchmarked nonnegative
portfolio $N^\delta$ is always a super\-martin\-gale. It is a
well-known fact that whenever a nonnegative supermartingale
reaches the value zero, it almost surely remains zero afterwards.
Based on this fundamental property of supermartingales,  no
company and no investor, with nonnegative total tradable wealth,
can generate wealth out of zero initial capital. This means that a
rather strong type of arbitrage, see \citeN{Platen02g}, is
automatically excluded in our market. Note however, free lunches
with vanishing risk in the sense of \citeN{DelbaenSc06}, or free
snacks and cheap thrills as described in \citeN{LoewensteinWi00},
may arise. This emphasizes that our financial market model is
rather general. In particular, as shown in \citeANP{Platen02g}
\citeyear{Platen02g,Platen04cc}, for the class of models under
consideration one does, in general, not have an equivalent risk
neutral probability measure. Therefore, the widely used risk
neutral pricing methodology may break down. This happens, for
instance, when the benchmarked savings account process $N^0$ forms
a {\em strict supermartingale\/} and not a martingale in a
complete market setting. For example, for realistic models where
such phenomenon arises we refer to \citeN{Sin98}, \citeN{Lewis00},
\citeN{LoewensteinWi00}, \citeN{Platen01a}, \citeN{MillerPl05i},
\citeN{FernholzKa05} and \citeN{PlatenHe06}.

Since risk neutral pricing is not available, we need a  general
consistent alternative for the pricing of contingent claims.  We
have already seen that benchmarked nonnegative portfolios are
supermartingales. It is clear that among those supermartingales
that replicate a given future benchmarked payoff, the
corresponding martingale provides the least expensive hedge. To
value claims consistently in a complete market, we generalize the
concept of {\em real world pricing}, as introduced in
\citeN{Platen02g} and \citeN{PlatenHe06}. It makes benchmarked
derivative prices to martingales by employing the GOP as numeraire
and forms in the resulting pricing formula conditional
expectations under the real world probability measure.

More precisely, let $\{H_t,t\geq0\}$ be an optional process, and
let us define the {\em payoff\/} $\Htau$, which matures at the
random time $\tau \in [0,\infty)$, as the nonnegative  random
variable with integrable benchmarked value, that is,
 \BE
 \label{ca4.2}
 \bE\l(\frac{\Htau}{X^\deltas_\tau} \r) < \infty.
 \end{equation}
When $\tau$ is a stopping time, then the payoff $H_\tau$ is called
a {\em contingent claim\/}. We define for $H_\tau$ its {\em real
world price\/} $V_\Htau(t)$ at time $t$  by the real world pricing
formula
 \BE
 \label{ca4.4}
 V_\Htau(t) = X^\deltas_t\, \bE \l(\frac{\Htau}{X^\deltas_\tau}\,\bigg|\,\cF_t\r)
 \end{equation}
for $t \in [0,\infty)$. This is a generalization of the real world
pricing concept described in \citeN{PlatenHe06}.

Note that by using real world pricing  for general derivatives,
the benchmarked unhedgeable part of a square integrable
benchmarked contingent claim has minimal variance since its
benchmarked current value is the least-squares projection of its
future benchmarked value.  On the other hand, replicable claims
can be hedged with minimal  costs. In the case when there exists a
minimal equivalent martingale measure in the sense of
\citeN{FollmerSc91} or \citeN{HofmannPlSc92}, the corresponding
risk neutral price is equivalent to the above real world price in
(\ref{ca4.4}).

\section{Honest Times}
 \setcA
 \label{sec:law4}
 \subsection{A simple characterization}

As previously shown, benchmarked nonnegative portfolios turn out
to be local martingales and, thus, supermartingales. If these are
modeled as transient jump diffusions, then there is always a last
time when these benchmarked securities reach their maximum. Such a
time is extremely interesting for an investor, since at this time
the portfolio reaches its largest value relative to the benchmark.
It would be beneficial if an investor could time the selling of a
security accordingly.  Unfortunately, the time when such a maximum
is reached is not a stopping time. However, already the knowledge
of the law of such time can provide the investor with precious
information.

Such random times are commonly called honest times, see Definition
\ref{def:law4.1} below. This class of random times is the most
studied one after stopping times. There are several
characterizations of honest times. One of these characterizations
is given in terms of nonnegative local martingales without
positive jumps, which vanish at infinity, and the last time they
reach their maximum. This corresponds well to the framework of our
financial market.

We first introduce an  abstract class of random times that
allows us to study a range of interesting problems related to the
above type of times.

\begin{definition}
 \label{def:law4.1} \quad Let $L$ be the end of an $\ucF$-optional
 set $\Gamma$, that is,
 \[ L = \sup\{t:\,(t,\omega) \in \Gamma\}. \]
 Then we call $L$ an {\em honest time}.
\end{definition}

An example for an honest time is the above mentioned time when a
benchmarked nonnegative portfolio without positive jumps reaches its last maximum.
Further examples will be given below.
\citeN{Azema72} associated with an honest time $L$ the
supermartingale
\[ Z^L_t = \bP(L > t\sb|\cF_t) = \mathbf{E}\l(\b1_{\{L>t\}} \sb|\cF_t\r) \]
and studied its properties. Note that this supermartingale plays a
key role in the enlargement of filtrations, as shown in
\citeN{Yor78}, \citeN{Jeulin80} or \citeN{JeulinYo85}. We now
provide a simple characterization of honest times and the
associated supermartingales, following the ideas of
\citeN{NikeghbaliYo06}.
\begin{definition}
 \label{def:law4.2} \quad We say that an $(\ucF,\mathbf{P})$-local
martingale $N=\{N_t, \tgo\}$ belongs to the class $(\cC_0)$, if it
is strictly positive, with no positive jumps and \\ $\lim_{t \to
\infty} N_t =0$, where $N_0=x>0$.
\end{definition}

Note that the benchmarked primary security account processes
$N^j$, $\,j \in \{0,1,\ldots,$ $d\}$, and all benchmarked
portfolios are local martingales. Consequently, all strictly
positive benchmarked portfolios with no negative jumps that vanish
at infinity  belong to this class. With the class $(\cC_0)$ we
cover benchmarked primary security accounts of realistic models.
Negative jumps of equities against the market as a whole are the
most important events we need to capture in financial modeling.
These are typically triggered by defaults or catastrophes.
Portfolios with no short sale constraint based on benchmarked
primary security accounts from the class $(\cC_0)$ are also from
$(\cC_0)$.  For notational convenience  we introduce for $N$ from
the class $(\cC_0)$ both future and past suprema processes:
 \[ \Sigma^t = \sup_{u\geq t} N_u \]
 and
 \[ \Sigma_t = \sup_{u\leq t} N_u. \]
The following variant of Doob's maximal inequality, see
\citeN{RevuzYo99}, also called {\em Doob's maximal identity}, has
far reaching consequences.

\begin{lem}[\citeN{NikeghbaliYo06}]
 \label{lem:law4.3} \ {\em (Doob's maximal identity)}\\
For any $a>0$ we have:
 \BE
 \label{law4.1}
 \bP(\Sigma_\infty >a) = \l(\frac xa \r) \wedge 1.
 \EE
Hence, $\frac{x}{\Sigma_\infty}$ is uniformly distributed on $(0,1)$.
Furthermore, for any stopping time $\tau$:
 \BE
 \label{law4.2}
 \bP(\Sigma^\tau >a \sb|\cF_\tau) = \l(\frac{N_\tau}{a} \r) \wedge 1.
 \EE
Here $\frac{N_\tau}{\Sigma^\tau}$ is a uniformly distributed random
variable on $(0,1)$, independent of $\cF_\tau$.
\end{lem}

\proof \quad Since this result is rather fundamental for our
analysis we indicate here its proof. Formula \eqref{law4.2} is a
consequence of equation \eqref{law4.1}, when applied to the
martingale $\tN=\{\tN_t = N_{\tau+t},\,t \geq 0\}$ and the
filtration $\tucF = (\cF_{\tau+t})_{t \geq 0}$. Formula
\eqref{law4.1} itself is obvious when $a\leq x$, and for $a>x$, it
is obtained by applying Doob's optional stopping theorem to the
local martingale $\hN=\{\hN_t = N_{t \wedge T_a}, \tgo\}$, where
$T_a = \inf\{u \geq 0:\,N_u>a\}$. \qBox

\begin{remark}
 \label{rem:law4.4} \quad The second part of
Lemma~\ref{lem:law4.3} with \eqref{law4.2}, is a remarkable
property that allows us to separate the distribution of
$\Sigma^\tau$ from the past, given the present. Without imposing
any Markovianity on the market dynamics one has the same
probabilistic characterization of the future supremum of a
benchmarked process, which only involves the simple uniform
distribution. From a finance point of view one can say that even
the most complex, possibly non Markovian,  jump diffusion dynamics
provide at any stopping time the same conditional probability
distribution for the maximum of a benchmarked security from
$(\cC_0)$ as is obtained, for instance, under the Black-Scholes
model.
\end{remark}
\begin{remark}
Under the minimal market model (MMM) (see \citeN{PlatenHe06}),
benchmarked primary security accounts are the inverse of time
transformed squared Bessel processes of dimension four, and thus
from the class $(\cC_0)$. For illustrations we show in
Figure~\ref{fig:lm1.1}, using realistic parameters, twenty
trajectories of these strict local martingales. One notes here
substantial movements upwards but overall a systematic decline,
consistent with the strict local martingale property. In
Figure~\ref{fig:lm1.2}, we display the running maximum $\Sigma_t$
for fifty paths $N_t$. Note that there seems to be no average
value identifiable if we would add more paths and would extend the
time horizon. We then show in Figure~\ref{fig:lm1.3} the inverse
of the running maxima. They seem to fit well a uniform
distribution on $(0,1)$, as is suggested by Lemma
\ref{lem:law4.3}.
\end{remark}

\Figgnunu{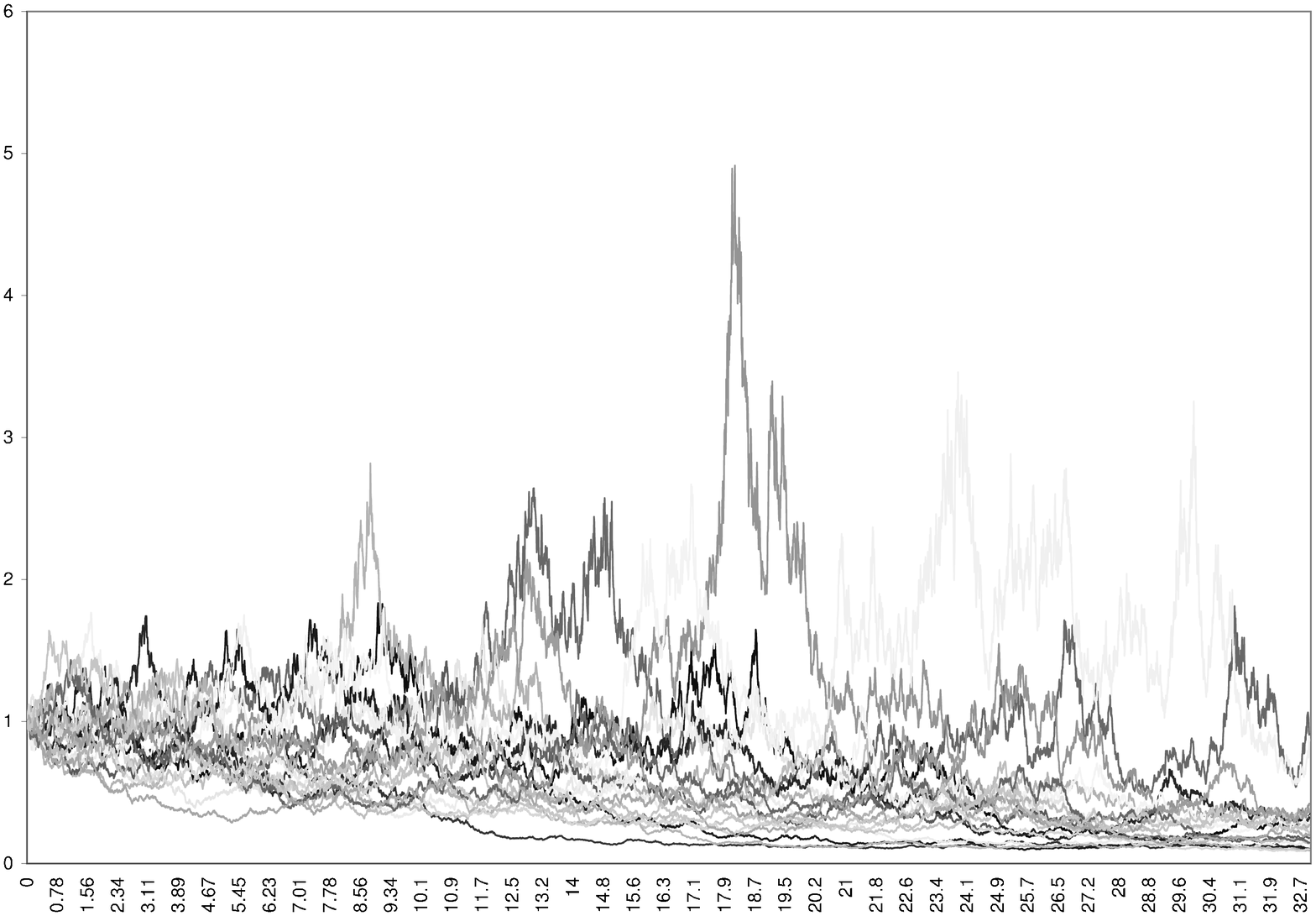}
 {Trajectories of $N_t$ under MMM}{fig:lm1.1}

 \Figgnunu{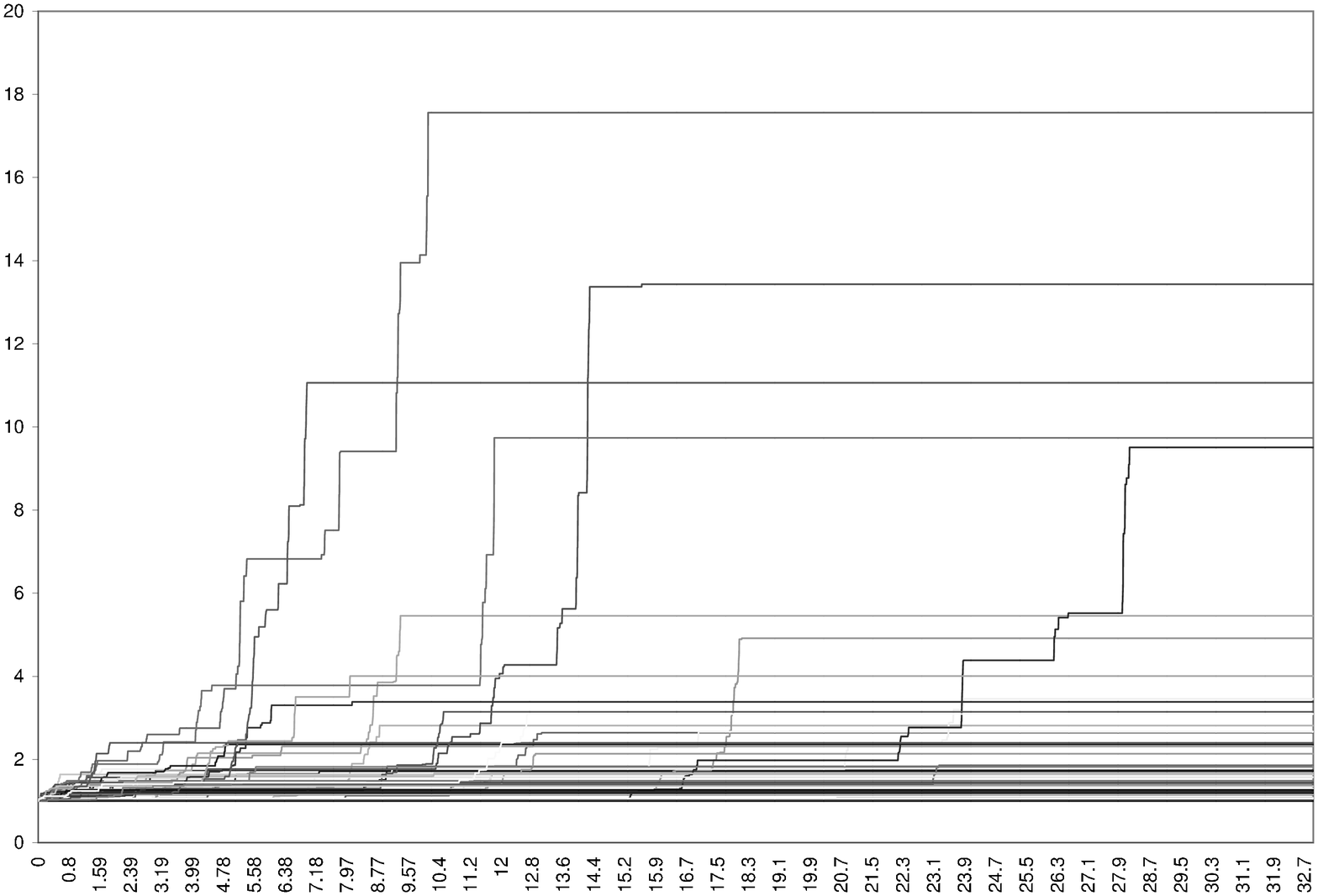}
 {Running maxima}{fig:lm1.2}

 \Figgnunu{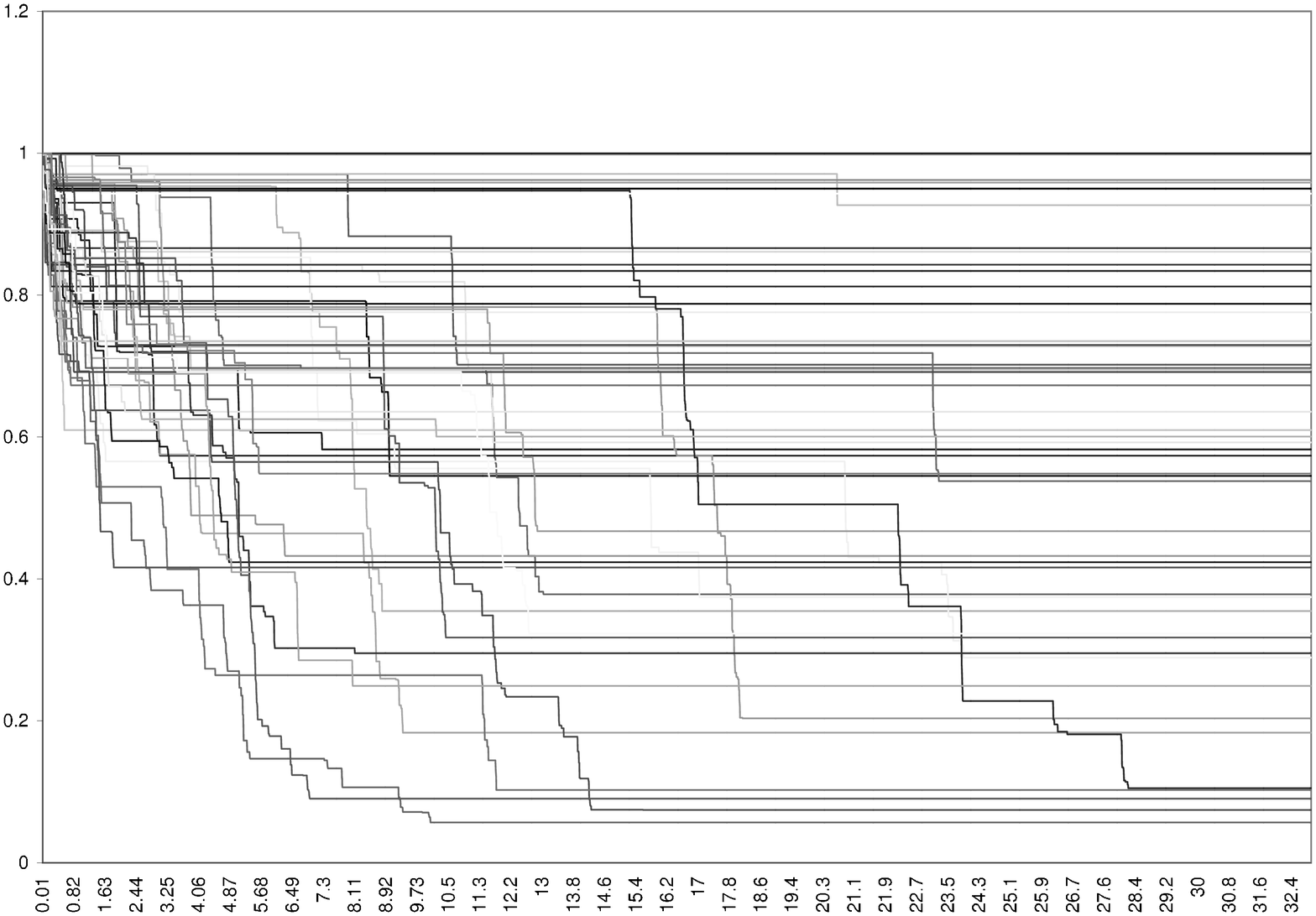}
 {Inverse of maxima}{fig:lm1.3}

The following proposition is rather interesting:

\begin{prop} [\citeN{NikeghbaliYo06}] \label{prop:law4.5} \quad
Let $N=\{N_t, \tgo\}$ be a local martingale, which belongs to the
class $(\cC_0)$, with $N_0=1$ and $\Sigma_t=\sup_{u \leq t}N_u$. When
we consider the honest time
 \BE
 \label{law4.3}
 g=\sup\{ \tgo:\,N_t=\Sigma_\infty\} = \sup\{\tgo:\,N_t=\Sigma_t\},
 \EE
we have  the formula
\[ Z_t = \bP(g > t\sb|\cF_t) = \frac{N_t}{\Sigma_t} \]
for all $\tgo$.
\end{prop}

This means that the ratio of the benchmarked security over its
maximum $\Sigma_t$ until time $t$, provides us with the
conditional probability that the time of the total maximum is
still ahead. When the benchmarked value $N_t$ of a security
substantially declines over time, compared to its maximum
$\Sigma_t$ until time $t$, then the conditional probability of the
time of the total maximum $\Sigma_\infty$ being still ahead is
becoming rather small equating the ratio $\frac{N_t}{\Sigma_t}$.
This important insight holds without any further modeling
assumption. In Figure~\ref{fig:lm1.4}, we illustrate this by
displaying trajectories of $N_t$, $\Sigma_t$ and
$\mathbf{P}(g>t|\cF_t)=\frac{N_t}{\Sigma_t}$.

\Figgnunu{Figure4.psg}
 {Trajectories of $N_t$, $\Sigma_t$ and $\mathbf{P}(g>t|\cF_t)=\frac{N_t}{\Sigma_t}$}{fig:lm1.4}

A direct application of the above
result together with the It\^o formula yields the representation
 \BE
 \label{law4.4}
 Z_t = \frac{N_t}{\Sigma_t} = 1+ \int^t_0 \frac{1}{\Sigma_s}\,dN_s
 -\ln(\Sigma_t).
 \EE
From the uniqueness of the Doob-Meyer decomposition it results
that $\ln(\Sigma_t)$ represents the increasing part  of $Z_t$
whilst $\int^t_0 \frac{1}{\Sigma_s}\,dN_s$ expresses its
martingale part. Since $Z_t$ is of the class $(\cD)$, $\int^t_0
\frac{1}{\Sigma_s}\,dN_s$ is a uniformly integrable martingale,
even in BMO. To keep the presentation as simple as possible, we
make the assumption for  the following two results,  that all
$\ucF$-local martingales are continuous.

\begin{cor}
 \label{cor:law4.7} \quad
Assume that all $\ucF$-local martingales are continuous. Let $g$
be defined in \eqref{law4.3}, then the quantity $ln(\Sigma_t)$ is
the dual predictable projection of the indicator $\b1_{\{g\leq
t\}}$, and for any positive predictable process $k =\{k_s,
s\geq0\}$ one has
 \BE
 \label{law4.5}
 \bE(k_g) = \bE\l(\int^\infty_0 k_s\,\frac{d\Sigma_s}{\Sigma_s}\r)
 =\bE\l(\int^\infty_0 k_s\,\frac{d\Sigma_s}{N_s}\r).
 \EE
Furthermore, the random time $g$ defined in \eqref{law4.3} is an
honest time and avoids any $\ucF$-stopping time $\tau$, that is,
$\bP(g=\tau)=0$.
 \end{cor}

This result is very useful for the quantitative analysis of
functionals that involve honest times. The following theorem
establishes a converse result
 to  Proposition~\ref{law4.5}.

\begin{theorem}[\citeN{NikeghbaliYo06}]
 \label{the:law4.8} \quad
Assume that all $\ucF$-local martingales are continuous, and let
$g$ be an honest time which avoids
 any $\ucF$-stopping time, then there exists a continuous,
 nonnegative local martingale $N=\{N_t, \tgo\}$ with $N_0=1$ and
 $\lim_{t \to \infty} N_t=0$, such that
 \BE
 \label{law4.6}
 Z_t = \bP(g>t \sb|\cF_t) = \mathbf{E}\l(\b1_{\{g>t\}}\sb|\cF_t\r)
 = \frac{N_t}{\Sigma_t}
 = \mathbf{E}\l(\ln(\Sigma_\infty) \sb|\cF_t\r) - \ln(\Sigma_t)
 \EE
for all $\tgo$.
\end{theorem}

This theorem states the remarkable fact that every honest time in
a continuous market is, in fact, the last time when a certain
nonnegative local martingale reaches its maximum. If the
continuous market is complete, it is the last time that a
nonnegative benchmarked portfolio reaches its maximum.
Furthermore, it follows by \eqref{law4.6} that the conditional
expectation for the logarithm of the maximum of a benchmarked
nonnegative portfolio is of the form
 \BE
 \label{law4.7}
 \mathbf{E}\l(\ln\l(\Sigma_\infty\r) \sb|\cF_t\r) = \frac{N_t}{\Sigma_t}+\ln\l(\Sigma_t\r)
 \EE
for $\tgo$. Note that the logarithm of $\Sigma_\infty$ is finite
and $R_t=\frac{N_t}{\Sigma_t}+\ln\l(\Sigma_t\r)$ is a martingale.
This provides  an investor with the possibility to identify
explicitly the conditional expectation of the logarithm of the
total maximum over the maximum until time $t$. It is important to
realize that the conditional expectation in (\ref{law4.7}) is
model independent, and is therefore very robust. More generally,
as is explained in the next subsection, one can even obtain the
conditional distributions of $\Sigma_\infty$. This allows to
evaluate general payoffs depending on $\Sigma_\infty$.

\subsection{Conditional distributions of $\Sigma_\infty$ and stochastic \\ integral representations}
In this subsection, we need not assume continuity of the local martingales.

The following relation can be rather useful in pricing derivative
payoffs since it applies to all benchmarked primary security
accounts and nonnegative portfolios from $(\cC_0)$. It gives
access to the conditional expectation of functions of
$\Sigma_\infty$, quantities all investors would like to know. Most
importantly, the results are again model independent. For sake of
completeness and to illustrate how a fundamental result such as
Doob's maximal identity leads easily to nontrivial statements, we
reproduce a proof of the following result:

\begin{prop}[\citeN{NikeghbaliYo06}]
For any Borel bounded or positive function $f$, and $N$ from $(\cC_0)$, we have:
\begin{eqnarray}
\mathbf{E}\left(f\left(\Sigma_{\infty}\right)|\cF_{t}\right) &=&
f\left(\Sigma_{t}\right)\left(1-\dfrac{N_{t}}{\Sigma_{t}}\right)+%
\int_{0}^{N_{t}/\Sigma_{t}}dyf\left(\dfrac{N_{t}}{y}\right)  \label{grosavecs} \\
&=& f\left(\Sigma_{t}\right)\left(1-\dfrac{N_{t}}{\Sigma_{t}}\right)+N_{t}%
\int_{\Sigma_{t}}^{\infty}dy\frac{f\left(y\right)}{y^{2}}.  \notag
\end{eqnarray}
\end{prop}
\proof \quad The proof is based on Doob's maximal identity; in the
following, $U$ is a random variable, which follows the standard
uniform law and which is independent of $\cF_{t}$.
\begin{eqnarray*}
\mathbf{E}\left(f\left(\Sigma_{\infty}\right)|\cF_{t}\right) &=& \mathbf{E%
}\left(f\left(\Sigma_{t}\vee S^{t}\right)|\cF_{t}\right) \\
&=& \mathbf{E}\left(f\left(\Sigma_{t}\right)\mathbf{1}_{\left\{\Sigma_{t}\geq
S^{t}\right\}}|\cF_{t}\right)+\mathbf{E}\left(f\left(S^{t}\right)%
\mathbf{1}_{\left\{\Sigma_{t}< S^{t}\right\}}|\cF_{t}\right) \\
&=& f\left(\Sigma_{t}\right)\mathbf{P}\left(\Sigma_{t}\geq S^{t}|\cF%
_{t}\right)+ \mathbf{E}\left(f\left(S^{t}\right)\mathbf{1}_{\left\{\Sigma_{t}<
S^{t}\right\}}|\cF_{t}\right) \\
&=& f\left(\Sigma_{t}\right)\mathbf{P}\left(U\leq \dfrac{N_{t}}{\Sigma_{t}}|\cF_{t}\right)+%
\mathbf{E}\left(f\left(\dfrac{N_{t}}{U}\right)\mathbf{1}_{\left\{U<\frac{%
N_{t}}{\Sigma_{t}}\right\}}|\cF_{t}\right) \\
&=& f\left(\Sigma_{t}\right)\left(1-\dfrac{N_{t}}{\Sigma_{t}}\right)+%
\int_{0}^{N_{t}/\Sigma_{t}}dxf\left(\dfrac{N_{t}}{x}\right).
\end{eqnarray*}
A straightforward change of variable in the last integral also gives:
\begin{equation*}
\mathbf{E}\left(f\left(\Sigma_{\infty}\right)|\cF_{t}\right)=f\left(\Sigma_{t}%
\right)\left(1-\dfrac{N_{t}}{\Sigma_{t}}\right)+N_{t}\int_{\Sigma_{t}}^{\infty}dy\frac{%
f\left(y\right)}{y^{2}},
\end{equation*}and this completes the proof. \qBox

Now, from  a financial point of view, it would be very useful to
obtain a representation of (\ref{grosavecs}) as a stochastic
integral. This would be very valuable for hedging purposes.
Remarkably, this can also be achieved without assuming continuity
of $N$ nor any predictable representation property for the
underlying filtration. Again, this result is universal since model
independent.  The next proposition  extends  the classical
Az\'{e}ma-Yor martingales:
\begin{prop}[\citeN{NikeghbaliYo06}]\label{azemayorgeneralisee}
Let $\left(N_{t}\right)_{t\geq 0}$ be from $(\cC_0)$, $f$ be a
locally bounded Borel function and define
$F\left(z\right)=\int_{0}^{z}dyf\left(y\right)$. Then, $X_{t}=
F\left(\Sigma_{t}\right)-f\left(\Sigma_{t}\right)\left(\Sigma_{t}-N_{t}\right)$ is
a local martingale and we have the representation:
\begin{equation}  \label{ayor}
F\left(\Sigma_{t}\right)-f\left(\Sigma_{t}\right)\left(\Sigma_{t}-N_{t}\right)=%
\int_{0}^{t}f\left(\Sigma_{s}\right)dN_{s}+F\left(\Sigma_{0}\right).
\end{equation}
\end{prop}
It is now easy to see that
$\mathbf{E}\left(f\left(\Sigma_{\infty}\right)|\cF_{t}\right)$ is
of the form (\ref{ayor}). Indeed:
\begin{eqnarray*}
\mathbf{E}\left(f\left(\Sigma_{\infty}\right)|\cF_{t}\right)&=&
f\left(\Sigma_{t}\right)\left(1-\dfrac{N_{t}}{\Sigma_{t}}\right)+N_{t}\int_{\Sigma_{t}}^{%
\infty}dy\frac{f\left(y\right)}{y^{2}} \\
&=&\Sigma_{t}\int_{\Sigma_{t}}^{\infty}dy\frac{f\left(y\right)}{y^{2}}
-\left(\Sigma_{t}-N_{t}\right)\left(\int_{\Sigma_{t}}^{\infty}dy\frac{f\left(y\right)}{%
y^{2}}-\dfrac{f\left(\Sigma_{t}\right)}{\Sigma_{t}}\right).
\end{eqnarray*}%
Hence,
\begin{equation*}
\mathbf{E}\left(f\left(\Sigma_{\infty}\right)|\cF_{t}\right)=H\left(1%
\right)+H\left(\Sigma_{t}\right)-h\left(\Sigma_{t}\right)\left(\Sigma_{t}-N_{t}\right),
\end{equation*}
with
\begin{equation*}
H\left(z\right)=z\int_{z}^{\infty}dy\frac{f\left(y\right)}{y^{2}},
\end{equation*}
and
\begin{equation}\label{troisdix}
h\left(z\right)=\int_{z}^{\infty}dy\frac{f\left(y\right)}{y^{2}}-\dfrac{%
f\left(z\right)}{z}=\int_{z}^{\infty}\frac{dy}{y^{2}}\left(f\left(y\right)-f\left(z\right)\right).
\end{equation}
Moreover, again from formula (\ref{ayor}), we obtain  the following
representation of $\mathbf{E}\left(f\left(\Sigma_{\infty}\right)|\cF%
_{t}\right)$ as a stochastic integral:
\begin{prop}
Let $N$ be from $(\cC_0)$ and $f$ be a Borel function such that
$\mathbf{E}\left(f\left(\Sigma_{\infty}\right)\right)<\infty$.
Then we have:
\begin{equation}  \label{represstoc}
\mathbf{E}\left(f\left(\Sigma_{\infty}\right)|\cF_{t}\right)=\mathbf{E}%
\left(f\left(\Sigma_{\infty}\right)\right)+\int_{0}^{t}h\left(\Sigma_{s}\right)dN_{s},
\end{equation}with $h$ as in (\ref{troisdix}).
\end{prop}

We obtained in (\ref{represstoc}) a martingale representation for
$\mathbf{E}\left(f\left(\Sigma_{\infty}\right)|\cF_{t}\right)$.
This can be exploited for hedging a payoff that involves the total
maximum $\Sigma_\infty$ of a benchmarked nonnegative portfolio
$N_t=\frac{X_t^\delta}{X_t^{\delta*}}$ from $(\cC_0)$. It turns
out that a call payoff $f(\Sigma_\infty)=(\Sigma_\infty-K)^+$
would have infinite expected value. However, a put payoff
$f(\Sigma_\infty)=(K-\Sigma_\infty)^+$ with benchmarked strike
$K\in(0,\infty)$ has according to the real world pricing formula
(\ref{ca4.4}) and (\ref{grosavecs}) at time $t$ the benchmarked
value
\[ \widehat{V}_t=\mathbf{E}\l(\l(K-\Sigma_\infty\r)^+|\cF_t\r)
=(K-\Sigma_t)^+(1-\frac{N_t}{\Sigma_t})
+N_t\int_{\Sigma_{t}}^{\infty}dy\frac{\left(K-y\right)^+}{y^{2}},\]
that is
\begin{equation}\label{vt}
 \widehat{V}_t=(K-\Sigma_t)^+(1-\frac{N_t}{\Sigma_t})
 +\mathbf{1}_{\Sigma_t<K}\;N_t(\frac{K}{\Sigma_t}-1-\ln
(\frac{K}{\Sigma_t})),
\end{equation}
or equivalently
\[\widehat{V}_t=\mathbf{1}_{\Sigma_t<K}\left(\left(K-\Sigma_t\right)-N_t\ln
(\frac{K}{\Sigma_t})\right).\]
Obviously, for $K\leq \Sigma_t$,
the value of the put is zero, since it gives the right but not the
obligation to receive the strike $K$ when paying
$\Sigma_\infty\geq\Sigma_t\geq N_t$. By (\ref{troisdix}) we can
determine the number of units
 \begin{eqnarray}
 \label{troisonze}
\nonumber   h(\Sigma_t) &=& \int_{\Sigma_t}^{\infty}\frac{(K-y)^+}{y^2}dy
 -\frac{(K-\Sigma_t)^+}{\Sigma_t} \\
   &=& -\mathbf{1}_{\Sigma_t<K}\;\ln (\frac{K}{\Sigma_t}),
\end{eqnarray}
that one has to hold in $X_t^\delta$ at the time $t$ to hedge the
payoff. The remainder of the wealth in the hedge portfolio should
be invested in the GOP $X_t^{\delta*}$. Note that when
$\Sigma_t=K$, the hedge portfolio collapses to zero and remains
there. The above put option can be used to protect against
downward moves of $N_t=\frac{X_t^\delta}{X_t^{\delta*}}$. For
instance, one can add the put to the security $N_t$ obtaining
$U_t=N_t+\widehat{V}_t$ as the benchmarked value of the resulting
portfolio. For the price of $\widehat{V}_0$, one purchases then at
time $t=0$ protection of some kind against downward moves of
$N_t$. Figure~\ref{fig:lm1.5} displays a corresponding scenario
where we show $N_t$, as it already appeared in
Figure~\ref{fig:lm1.4}, and $U_t$ when $K$ was set equal to $2.5$.
In this scenario the strike $K$ was not reached by $\Sigma_t$
during the period displayed. One notes that the value $U_t$ stays
always above $N_t$, which it should by construction. However, we
also see that $U_t$ does practically not fall much below a level
of about $0.7$, which is close to the final value of $K-\Sigma_t$.
This illustrates the kind of protection that $U_t$ is giving
against downward moves of $N_t$. For periods when $N_t$  comes
closer to $K$, $U_t$ goes up. However, when $N_t$  falls quite
drastically, $U_t$ demonstrates its put feature. Overall it
appears that $U_t$ benefits from extreme upward  moves of $N_t$.
In the long term, the systematic downward movement of $N_t$ will
be in $ \widehat{V}_t$ asymptotically limited to
$K-\Sigma_\infty$.

\Figgnunu{Figure5.psg}
{Trajectories of $N_t$ and $U_t$}{fig:lm1.5}

 This simple example indicates that there  exist many  ways of creating new financial products or managing risk  using the above results on honest times. What is most striking for the above pricing and hedging results, involving payoffs based on $\Sigma_\infty$, is the fact that these are model independent and therefore very robust. This is a property that makes the methodology very attractive for areas where modeling risk over a long period of time has been of much concern, as it is in pension fund management and insurance. Furthermore, it is not  that one wants to actually receive the payoff $f(\Sigma_\infty)$, it is more that one is aiming for it. In this manner the use of honest times creates a new perspective for risk management.
\section{Law of an Honest Time}
 \setcA
 \label{sec:law5}
After having invested over some time period in a security, it is a
successful strategy to sell such security when its benchmarked
value $N_t$ comes close to its total maximum $\Sigma_\infty$. The  problem with such buy low sell high strategy is that the investor cannot decide at a given time if
 the total maximum has occurred or not
because this  arises at an honest time which is not a stopping time. For an investor it is
therefore of interest to know at least the law of the hidden time of the
total maximum of a benchmarked security. This allows her or him to judge whether it is realistic to hope, over a given time frame, to reach the total maximum. We will  give below a
formula that will be useful for the explicit computation of the
law of an honest time.

\begin{prop}
 \label{prop:law5.1} \quad
Under the assumptions of Theorem~\ref{the:law4.8} the law of the
honest time $g$ is given as
 \BE
 \label{law5.1}
 \bP(g \leq t) = \bE(\ln(\Sigma_t)).
 \EE
\end{prop}

\proof \quad We have by \eqref{law4.4} and \eqref{law4.6}
 \[ P(g>t\sb|\cF_t) = \frac{N_t}{\Sigma_t} =
 1+\int^t_0 \frac{1}{\Sigma_s}\,dN_s - \ln(\Sigma_t). \]
Taking expectation on both sides and exploiting the fact that
$\int^t_0 \frac{1}{\Sigma_s}\,dN_s$ is a uniformly integrable
martingale yields \eqref{law5.1}. \qBox \\

Although formula \eqref{law5.1} seems to be rather simple it
still requires the knowledge of the mean of $\ln(\Sigma_t)$. Obviously, the law of an honest time is no longer model independent. Numerical
methods, in particular Monte Carlo methods, as described for instance in
\citeN{KloedenPl92}, can be very useful in such computations.
However, it is of great advantage in the study of an honest time
if one can derive an explicit analytic formula for its law. Below we provide a
theorem that can be  useful when aiming to compute explicitly the law
of an honest time.

\begin{theorem}
 \label{the:law5.2} \quad
Under the assumptions of Theorem~\ref{the:law4.8} define $\tau_a =
\inf\{t:\,N_t >a\}$ for $a \geq 1$. Then for any bounded or
positive Borel function $f$, we have
 \BE
 \label{law5.2}
 \bE(f(g)) = \int^\infty_1 \bE\l(f(\tau_a)\,
 \b1_{\{\tau_a<\infty\}}\r) \frac{da}{a}.
 \EE
In particular, the Laplace transform of the law of $g$ is obtained
as
 \BE
 \label{law5.3}
 \bE(\exp\{-\lambda\,g\}) = \int^\infty_1 \bE\l(\exp\{-\lambda\,\tau_a\}\r)
  \frac{da}{a}
 \EE
for $\lambda >0$.
\end{theorem}

\proof \quad From Corollary~\ref{cor:law4.7}, or equally by differentiating formula (\ref{law5.1}), it follows for any
Borel bounded function $f$
 \begin{eqnarray*}
 \bE(f(g)) \sq \bE \l(\int^\infty_0 f(s)\,\frac{d\Sigma_s}{\Sigma_s}\r)
 = \bE\l(\int^{\Sigma_\infty}_1 f(\tau_a)\,\frac{da}{a}\r) \n2
 \sq \int^\infty_1 \bE\l( f(\tau_a)\,\b1_{\{\tau_a < \infty\}} \r)
 \frac{da}{a}. \qquad \qquad \Boxf
 \end{eqnarray*}
 \vso

This result allows us to derive some explicit examples for the law
of an honest time when the underlying nonnegative local martingale
$N$ belongs to some class of well-known diffusions. We begin with
the standard asset price model in finance, the Black-Scholes
model. We set
 \BE
 \label{law5.3a}
 N_t = \exp\{2\,\sigma\,W_t -2\,\sigma^2\,t\},
 \EE
which follows a  geometric Brownian motion for $\tgo$. Here
$W=\{W_t, \tgo\}$ denotes a standard Wiener process under the real
world probability measure $\mathbf{P}$ and we assume $\sigma
>0$. The honest time considered here, that is the time of the total maximum of $N_t$, is then given as
 \[ g=\sup \left\{t\geq0:\, (W_t-\sigma\,t) = \sup_{s \geq 0} (W_s-\sigma\,s)
 \right\}. \]

\begin{prop}
 \label{prop:law5.3} \quad The law of $g$ is characterized by its Laplace
 transform
 \BE
 \label{law5.4}
 \bE(\exp\{-\lambda\,g\}) =
 \frac{2}{1+\sqrt{1+\frac{2\,\lambda}{\sigma^2}}}
 \EE
for $\lambda \geq 0$.
\end{prop}

\proof \quad We can use \eqref{law5.3} to compute the law of $g$.
For this we will use the Laplace transform of $\tau_a =
\inf\{t:\,N_t>a\}$, which is obtained in \citeN{Williams74} and
also given in \citeN{BorodinSa96} as
 \BE
 \label{law5.5'}
 \bE(\exp\{-\lambda\,\tau_a\}) =
 \l(\frac{1}{a}\r)^{\sqrt{\frac 14 +\frac{\lambda}{2\,\sigma^2}}+\frac 12}
 \EE
for $a>1$ and $\lambda \geq 0$. Substituting formula
\eqref{law5.5'} into \eqref{law5.3} yields
 \begin{eqnarray*}
 \bE(\exp\{-\lambda\,g\}) \sq
 \int^\infty_1 \l(\frac{1}{a}\r)^{\sqrt{\frac 14 +\frac{\lambda}{2\,\sigma^2}}+\frac 12}
 \,\frac{da}{a}
 =  \int^\infty_0 e^{-u \l(\sqrt{\frac 14 +\frac{\lambda}{2\,\sigma^2}}+\frac 12\r)}
 \,du \n2
 \sq \frac{2}{1+\sqrt{1+\frac{2\,\lambda}{\sigma^2}}}. \qquad \qquad \Boxf
 \end{eqnarray*}
 \vso

\def\te{\tilde{e}}
 \def\law{{\hbox{\rm \scriptsize law}}}
\def\lDt{{\langle D \rangle_t}}
\def\lDs{{\langle D \rangle_s}}
\def\lD{{\langle D \rangle}}

\begin{remark}
 \label{rem:law5.5}: \quad It is interesting to note that
by \eqref{law5.4} the honest time $g$ has the same law as the
first hitting time of a level twice the value of an independent
exponential random variable $\te$ by a Brownian motion with drift,
that is,
\begin{equation}\label{itstranslation}
 g \stackrel \law = \frac{1}{\sigma^2}\,T_{ \te/2}
\end{equation}
with $T_a = \inf\{t:\,\tW_t +t=a\}$, where $\tW_t$ follows a
standard Brownian motion. Indeed, formula \eqref{law5.4}, or its translation \eqref{itstranslation}, can be seen as a particular case of the path decomposition of a transient diffusion, here $(\tW_t-t)_{t\geq0}$, as presented in \citeN{Jeulin80}, p.112, Proposition (6,29). More precisely, from this proposition (or from Doob's maximal identity), it follows that $$\sup_{t\geq0}(\tW_t-t)\stackrel \law =\te/2.$$ From the same proposition, we also learn that conditionally on $\sup_{t\geq0}(\tW_t-t)=a$, the process $(\tW_t-t;\;t\leq g)$ is distributed as $(\tW_t+t;\;t\leq T_a)$, since $(\tW_t+t)$ is the Doob $h$-transform of $(\tW-t)$ with $h(x)=\exp(2x)$.
 \end{remark}

The Black-Scholes model is still a simple model. Therefore, the
following result is of interest to give access to a much richer
class of models. By the Dubins-Schwarz theorem, see
\citeN{RevuzYo99}, we have the following characterization of local
martingales of the class $(\cC_0)$, which reduces the problem of
finding the law of an honest time $g$ for a general model to that
of a geometric Brownian motion after some time change.

\begin{prop}
 \label{prop:law5.6} \quad
Under the assumptions of Theorem~\ref{the:law4.8} let $g$ be an
honest time. Then there exists a unique local martingale $D =
\{D_t,\,\tgo\}$ with $\langle D \rangle_\infty = \infty$ a.s.\ and
$D_t = \int^t_0 \frac{dN_u}{N_u} = W_\lDt$, where $W$ is an
$(\cF_{\langle D\rangle^{-1}_u})$-Brownian motion, such that
 \[ g=\sup\left\{t:\,W_\lDt - \frac 12 \lDt = \sup_{s \geq 0}
 \l(W_\lDs - \frac 12 \lDs\r)\right\}. \]
\end{prop}

\proof \quad From \citeN{RevuzYo99} it follows that there exists a
local martingale $D$ such that $\lD_\infty = \infty$ and
$N_t=\exp\{D_t-\frac 12 \,\lDt\}$. Moreover, the local martingale
$D$ is unique and $D_t=\int^t_0 \frac{dN_u}{N_u}$. From the
Dubins-Schwarz theorem there exists then an $(\cF_{\langle
D\rangle^{-1}_u})$-Brownian motion $W=\{W_u,\;u\geq0\}$ in
$\lDt$-time such that $D_t=W_\lDt$. If we denote by $\lD_u^{-1}$,
the generalized inverse of $\lD_t$ defined by
 \[ \lD_u^{-1} = \inf\{t \geq 0:\,\lDt >u\}, \]
then we can define the honest time
 \[ L= \sup \left\{\tgo:\,W_u - \frac 12\,u =
 \sup_{s \geq 0} \l(W_s - \frac 12\,s\r)\right\}. \]
Consequently, $g=\lD_L^{-1}$ is also given by
 \[ g=\sup\left\{t:\,W_\lDt - \frac 12 \lDt = \sup_{s \geq 0}
 \l(W_\lDs - \frac 12 \lDs\r)\right\}. \qquad \Box \] %\qBox \\

Squared Bessel processes play an essential role in various
financial models. This includes, for instance, the constant
elasticity of variance model, see \citeN{Cox75}; the affine
models, see \citeN{DuffieKa94}; and the minimal market model, see
\citeANP{Platen02g} \citeyear{Platen01a,Platen02g}. To study
honest times in some of these models let $R^2=\{R^2_t, \tgo\}$
denote a squared Bessel process of dimension $\delta >2$. In this
case $R^2$ is transient, see \citeN{RevuzYo99}. Furthermore, for
any squared Bessel process with index $\nu =\frac{\delta}{2}-1>0$
the process $N=\{N_t, \tgo\}$ with
 \BE
 \label{law5.7}
 N_t = \l(\frac{R^2_0}{R^2_t}\r)^\nu
 \EE
is a nonnegative, strict local martingale and from the class
$(\cC_0)$ with $N_0=1$. By application of
Proposition~\ref{prop:law4.5} one obtains for the honest time
$g=\sup\{\tgo:\,R^2_t=I_t\}$ with $I_t=\inf_{s \leq t} R^2_s$ the
conditional probability
 \BE
 \label{law5.8}
 \mathbf{P}(g>t \sb|\cF_t) = \l(\frac{I_t}{R^2_t}\r)^\nu
 \EE
for all $\tgo$. For illustration, we show in
Figure~\ref{fig:lm1.7}, in the case of dimension $\delta=4$, some
simulated paths of $R_t^2$, $I_t$ and  the evolution of the
conditional probability (\ref{law5.8}).

\Figgnunu{Figure7.psg}
 {Trajectories of $\mathbf{P}(g>t|\cF_t)$, $R^2_t$ and $I_t$}{fig:lm1.7}

Moreover, by Doob's maximal identity \eqref{law4.1} the random
limit $\ 1/I_\infty$ is uniformly distributed on $(0,1)$ for the
case of dimension $\delta=4$. This is an interesting observation
for the rather realistic  minimal market model, where such
dynamics arise. Figure~\ref{fig:lm1.3} displays the running values
of $1/I_t$ for $50$ such scenarios.

\begin{prop}
 \label{prop:law5.8} \quad The Laplace transform of the
 honest time $g$ given in (\ref{law5.8}) is for $\lambda >0$ of the form
 \BE
 \label{law5.9}
 \bE(\exp\{-\lambda\,g\}) =
 \frac{2\,\nu\,K_\nu(\sqrt{2\,\lambda\,x)}}{(2\,\lambda\,x)^{\frac
 {\nu}{2}}} \ \int^{\sqrt{2\lambda x}}_0
 \frac{u^{\nu-1}}{K_\nu(u)}\,du,
 \EE
where $R^2_0=x$ and $K_\nu(\cdot)$ is the modified Bessel function
of the second kind, see \citeN{BorodinSa96}.
 \end{prop}

\proof \quad We first recall the Laplace transform of the random
variable $\tau_a = \inf\{\tgo:\,N_t=a\} =
\inf\{\tgo:\,R^2_t=\frac{x}{a^{\frac{1}{\nu}}}\}$, $a \geq x^\nu$,
from \citeN{Kent78} and \citeN{BorodinSa96} in the form
 \BE
 \label{law5.10}
 \bE(\exp\{-\lambda\,\tau_a\}) =\frac{K_\nu(\sqrt{2\,\lambda\,x})}
 {a\,K_\nu\!\l(\frac{\sqrt{2\,\lambda\,x}}{a^{\frac{1}{\nu}}}\r)}
 \EE
for $\lambda >0$. A combination of \eqref{law5.10} and
\eqref{law5.3} gives \eqref{law5.9}. \qBox \\

In the special case of dimension $\delta=4$, as it arises for the
stylized minimal market model in \citeANP{Platen02g}
\citeyear{Platen01a,Platen02g}, we have $\nu=1$ and it follows
that
 \[  \bE(\exp\{-\lambda\,g\}) =\frac{2\,K_1(\sqrt{2\,\lambda\,x})}
 {\sqrt{2\,\lambda\,x}} \ \int^{\sqrt{2\lambda x}}_0 \frac{1}{K_1(u)}\,
 du. \]
Another interesting special case is obtained for the squared
Bessel process of dimension $\delta=3$, where we are able to
provide the following explicit formula for the density.

\begin{cor}
 \label{cor:law5.9}  For dimension $\delta=3$ the law of the
honest time $g$ given in (\ref{law5.8}) has the density
\begin{equation}\label{law5.11}
 p(t) = \dfrac{1}{\sqrt{2\pi x t}}\left(1-\exp\left\{\frac{-x}{2t}\right\}\right),
\end{equation}
where $R^2_0 =x>0$ and $t\geq0$.
 \end{cor}

\proof \quad For the squared Bessel process of dimension
$\delta=3$ we have $\nu=\frac 12$ and from \eqref{law5.10}
 \[  \bE(\exp\{-\lambda\,g\}) = \frac{2}{\sqrt{2\,\lambda\,x}}
 \,\exp\left\{-\frac{\sqrt{2\,\lambda\,x}}{2}\right\} \
 \sinh\l(\frac{\sqrt{2\,\lambda\,x}}{2}\r).
 \]
The linearity of the Laplace transform and a close look at a table
of inverse Laplace transforms, see for instance
\citeN{BorodinSa96}, then yields \eqref{law5.11}.
\qBox \\

Note that the above density \eqref{law5.11} of the honest time is
dependent on the initial level of the squared Bessel process. Such
dependence was not observed in the case of geometric Brownian
motion.

Consider for the moment $1/N_t=\frac{X_t^{\delta*}}{X_t^0}$, the
inverse of a benchmarked savings account, which is  the discounted
GOP. Then the honest time in (\ref{law5.8}) describes the ideal
time to invest in the GOP funds that were held in the savings
account, waiting for investment in the best performing portfolio,
the GOP. The formulae (\ref{law5.10}) and (\ref{law5.11}) describe
the laws of this time for Bessel models. This information could be
used by the investor for the optimal timing of investment
decisions.

To have an even richer class of models for benchmarked securities
than those just discussed, we consider the general case of a
transient diffusion $Y=\{Y_t, \tgo\}$. It generates a local
martingale $N$ in the class $(\cC_0)$ via the ratio $N_t = \frac{
s(Y_t)}{s(x)}$, $\,\tgo$, $Y_0=x>0$. Here $s(\cdot)$ is the
differentiable scale function of $Y$, see \citeN{BorodinSa96},
which we can choose such that $s(0)=-\infty$ and $s(\infty)=0$.
Then we have by Proposition~\ref{prop:law4.5} again
 \[ \bP(g>t\sb|\cF_t) = \frac{ s(Y_t)}{s(Z_t)}, \]
where $Z_t = \inf_{s \leq t} Y_s$ and $g$ is defined as
$g=\sup\{\tgo:\,Y_t=Z_t\}$. The law of the honest time $g$ can
then be characterized as follows.

\begin{prop}
 \label{prop:law5.10} \quad The Laplace transform of the above
 honest time $g$ is for $\lambda >0$ of the form
 \BE
 \label{law5.12}
\bE(\exp\{-\lambda\,g\}) = - \int^x_0 \frac{s'(u)}{s(u)} \
 \frac{\varphi_\lambda(x)}{\varphi_\lambda(u)}\,du.
 \EE
Here $\varphi_\lambda(\cdot)$ is a continuous solution of the
equation
 \BE
 \label{law5.13}
 G \,\varphi_\lambda(y) = \lambda\, \varphi_\lambda(y),
 \EE
with $G$ denoting the infinitesimal generator of the diffusion
$Y$.
 \end{prop}

The function $\varphi_\lambda(\cdot)$ is characterized as the
unique (up to a multiplicative constant) solution of
\eqref{law5.13} by demanding that $\varphi_\lambda(\cdot)$ is
decreasing and satisfies some appropriate boundary conditions. The
reader is referred to \citeN{PitmanYo99} for further details on
the function $\varphi_\lambda(\cdot)$ and its relation to hitting
times.

{\bf Proof of Proposition~\ref{prop:law5.10}:} \quad Let us
consider the hitting time
\[ \tau_z = \inf\left\{\tgo: \,\frac{s(Y_t)}{s(x)}=a \right\}
 = \inf\{\tgo:\, Y_t=s^{-1}(a\,s(x))\} \]
for $a \geq 1$ and $z:= s^{-1}(a\,s(x)) \leq x$. The Laplace
transform of $\tau_z$ follows by \citeN{PitmanYo99} and
\citeN{BorodinSa96} in the form
 \BE
 \label{law5.14}
\bE(\exp\{-\lambda\,\tau_z\}) =
 \frac{\varphi_\lambda(x)}{\varphi_\lambda(z)}.
 \EE
It suffices to substitute $z=s^{-1}(a\,s(x))$ in \eqref{law5.14}
and then apply the resulting expression in \eqref{law5.3}. \qBox

Let us conclude this paper with the following remarks.

In this paper, we have not provided any specific rule on how to
use the knowledge of the law of an honest time $g$ for trading
strategies. One way would be to look for the closest stopping time
to $g$, with respect to some suitable distance. Such problems have
already been solved by \citeN{ToitPe07} in the case of the
Brownian motion with drift. In a forthcoming work, we will address
this problem both analytically and numerically, for various
models.

For sake of clarity, we have not included any discussion about
situations where the nonnegative local martingale $N$ does not
converge anymore to $0$, but rather to some random variable
$N_\infty$. Such results would be of interest in the case when the
trading period is $[0,T]$, for $T$ representing some fixed
deterministic time or some stopping time. In such situations, the
computations are more involved and are the topic of current
research. It is also obvious that the above presented results are
useful in the study of insider trading, see
\citeN{AmendingerImSc98} and \citeN{GrorudPo98} or
\citeN{Imkeller02}. Forthcoming work will address this issue by
using honest times.

% \subsection*{Acknowledgement}

%\bibliographystyle{\dir chicago}
% \bibliography{\dir my}

\end{document}